\documentclass[prc,aps,onecolumn,showpacs]{revtex4}
\begin{document}
\title{How Robust is the Froissart Bound? }

\author{Ya.~I.~Azimov$^{\,a}$}

\affiliation{$^a$Petersburg Nuclear Physics Institute,\\
                 St.~Petersburg, 188300, Russia\\}

\bigskip

\begin{abstract}

Proof of the Froissart theorem is reconsidered in a different
way to extract its necessary conditions. Two physical inputs,
unitarity and absence of massless intermediate hadrons, are
indisputable. Also important are mathematical properties of the
Legendre functions. Assumptions on dispersion relations, single
or double, appear to be excessive. Instead, one should make
assumptions on possible high-energy asymptotics of the
amplitude in nonphysical configurations, which have today no
firm basis. Asymptotics for the physical amplitude always
appear essentially softer than for the nonphysical one. Froissart's
paper explicitly assumed the hypothesis of power behavior and
obtained asymptotic bound for total cross sections $\sim
\log^2{(s/s_0)}$ with some constant $s_0$. Our bounds are
slightly stronger than original Froissart ones. They show
that the scale $s_0$ should itself slowly grow with $s$. Under
different assumptions about asymptotic behavior of nonphysical
amplitudes, the total cross section could grow even faster than
$\log^2{s}$. The problem of correct asymptotics might be
clarified by precise measurements at the LHC and higher energies.

\end{abstract}


\maketitle

\section{Introduction}

One of the cornerstones for the present strong interaction physics
is the Froissart theorem~\cite{frois}. It declares that the
total cross sections of any two-hadron scattering cannot grow
with energy faster than $(\log{s})^2$. The expectation agrees
with current experimental data~\cite{block}. However, the same
data can also be fitted so to have an asymptotical increase as
$s^\delta$ with $\delta\sim0.08$~\cite{lands}. Of course, some
authors suggest that the power behavior is only temporal and
will change at some very high energies. If, however, such a fit
continued infinitely, it would violate the canonical Froissart
bound. Thus, experimental validity of this bound stays an open
question. Measurements at LHC may help to clarify it.

Theoretically, the Froissart theorem was initially proved
for $2\to2$ amplitudes~\cite{frois} in the framework of the
double-dispersion representation (the Mandelstam
representation)~\cite{mand} with a finite number of subtractions.
Such representation is true in the nonrelativistic quantum
mechanics with Yukawa-type potentials~\cite{regge}. However,
it has never been proven mathematically for any relativistic
amplitude.

Froissart's bound for total cross-sections was reproduced by
Martin~\cite{mart} without any dispersion representations, and
even without considering amplitudes. In his proof, the
unitarity condition and absence of angular singularities in the
physical region were applied only to the absorptive part of an
elastic amplitude. He explicitly assumed also that the
absorptive part in the physical region may grow with energy not
faster than some power of $s$.

Later, however, Martin returned to the investigation of the
amplitude as a whole~\cite{martnc} (sure, the amplitude
contains more information than its absorptive part). Using the
basic principles of axiomatic local field theory, he extended
the analyticity domain of the scattering amplitude so that it
reveals at least part of the Mandelstam cuts. This has appeared
to be sufficient for confirming Froissart's results for the
amplitude. Moreover, such an approach has allowed many new
results to be obtained (\textit{e.g.}, Ref.\cite{lukmart}; see
also the recent paper~\cite{wmrs} and references therein). It
was further suggested, that a more accurate account for
unitarity could even improve existing bounds~\cite{mart08}.

There is, however, an interesting problem in the Martin approach.
Axiomatics of the quantum field theory suggest, in particular,
that the theory is constructed from local quantized fields,
which are related with isolated (quasifree) asymptotic
\textit{in} and \textit{out} states.

Meanwhile, Froissart-Martin boundaries are usually applied to
hadron processes. It is common belief now that the hadron
interactions are underlaid by the quantum chromodynamics (QCD).
However, QCD is hard to consider an axiomatic theory.
Indeed, it deals with quark and gluon fields, which are local,
but (because of confinement) cannot have isolated one-quark
and/or one-gluon states.

On the other hand, hadrons, consisting of quarks and gluons,
cannot be pointlike. Therefore, ``effective QCD'', dealing
directly with hadrons, should contain some nonlocality
(imagine description of atoms without explicit use of charged
nuclei and electrons).

Thus, application of the ideas and methods of axiomatic local
field theory to hadron properties might look dubious, as well as
application of dispersion relations. That is why we reconsider
here the derivation of Froissart's results, without any
hypotheses on double-dispersion, or even single-dispersion,
representations, or axioms of quantum field theory. In this way
we clarify the origin and necessary inputs for the Froissart
bound and can discuss their reliability. The present approach
allows us also to use stricter inequalities for the Legendre
functions and, thus, slightly improve the original Froissart
bound for total cross sections or other observables.

The presentation here goes as follows. In Sec. II we
demonstrate that only a finite number of partial-wave amplitudes
are essential at each particular energy. Section III shows
construction of bounds for amplitudes in different
configurations, either studied before or not, in terms of the
number of essential partial waves.

\section{Modified Froissart derivation}

To begin with, we go along Froissart's lines as close
as possible. Just as in Ref.\cite{frois}, we consider
a reaction of the type
\begin{equation}
a+b\to c+d
\label{react}
\end{equation}
among scalar particles. We shall assume that all masses
are equal to $m$ as we deal only with asymptotic properties,
where the difference between the masses is expected to be
negligible. We introduce the familiar Mandelstam variables
$s=(p_a+p_b)^2$, $\,t=(p_c-p_a)^2$ and $u=(p_d-p_a)^2$. Then
$s+t+u=4m^2$. Evidently, $s$ is the c.m. energy squared for
the reaction~(\ref{react}), which is called the $s$-channel
reaction. A similar role is played by $t$ and $u$ for two
cross-channel reactions
\begin{equation}
a+\bar{c}\to \bar{b}+d\,,~~~
\bar{c}+b\to \bar{a}+d\,,    \label{cross}
\end{equation}
respectively the $t$-channel and $u$-channel ones. For the
$s$-channel reaction, $t$ and $u$ are momentum transfers
squared, related to the reaction angle $\theta_s$ as
\begin{equation}
\cos\theta_s=1+(t/2q_s^2)=-1-(u/2q_s^2)\,,
\label{angle}
\end{equation}
where $q_s^2=(s-4m^2)/4$, $\,q_s$ being the $s$-channel c.m.
momentum. The $s$-channel physical region is given by
\[q_s^2 > 0\,,~~~|\cos\theta_s|\le 1\,;~~~~
\textrm{or}~~~~s>4m^2\,,~~~t\le 0\,,~~~u\le0\,. \]
Let us relate the reaction amplitude with the partial-wave
amplitudes (we use the same normalization and relations for
the amplitudes as Froissart~\cite{frois}):
\begin{equation}
A(s,\cos\theta_s)=\frac{\sqrt{s}}{\pi q_s}
\sum_{l=0}^{\infty}\,a_l(s)\,(2l+1)\,P_l(\cos\theta_s)\,,
\label{pw-ser}
\end{equation}
\begin{equation}
a_l(s)=\frac{\pi q_s}{2\sqrt{s}}\int_{-1}^{\,+1}
A(s,\cos\theta_s^{\,'})\,P_l(\cos\theta_s^{\,'})\,
d(\cos\theta_s^{\,'})\,.
 \label{pw-am}
\end{equation}
For any physical (integer) $l\,,$ $|a_l|$ is bounded by one,
$a_l$ being an element of the unitary $S$-matrix.

At the next step, Froissart~\cite{frois} uses the momentum
transfer dispersion relation for the amplitude at fixed $s$
to show that $a_l$ exponentially decreases at large $l\,$.
We go another way, without any dispersion relations. Note,
first of all, that $P_l(z)$ are analytical functions of $z$
in the whole $z$-plane and
$$|P_l(z)|<|P_l(\pm1)|=1~~~\textrm{at}~~-1<z<1\,.$$
If the series (\ref{pw-ser}) for the amplitude and for its
angular derivative are convergent in the end points
$\cos\theta_s=\pm1$, then they are convergent also for any
physical $\cos\theta_s$, and the amplitude $A(s,
\cos\theta_s)$ is an analytical function of $\cos\theta_s$
inside the whole physical region (and nearby).

As is well known, in addition to $P_l(z)\,$, the Legendre
functions of the 1st kind, there exist also the Legendre
functions of the 2nd kind, $Q_l(z)\,$. They are also analytical
functions of $z$, but have branch points. At integer values of
$l$, the 1st Riemann sheet contains only one cut, between $-1$
and $+1$. The jump over this cut is
\begin{equation}
\frac1{2i}\left[Q_l(x+i\epsilon)-Q_l(x-i\epsilon)\right]=
-\frac{\pi}2\,P_l(x),~~~~-1<x<+1\,.
\label{jump}
\end{equation}
Therefore, expression (\ref{pw-am}) may be rewritten as
\begin{equation}
a_l(s)=\frac{q_s}{2i\sqrt{s}}\oint A(s, z')\,Q_l(z')\,dz'\,,
\label{cint}
\end{equation}
where integration runs along the closed contour going
counterclockwise around the cut of $Q_l(z)$ between $-1$ and $+1$.
Such a contour crosses the real $z$-axis both at $z>+1$,
and at $z<-1$. Let us define $$z'=\cosh(\alpha+i\,\varphi')$$
with real $\alpha,\,\varphi'$. We can choose $\alpha\ge0\,$
and construct the contour in Eq.(\ref{cint}) so to have
a constant value of $\alpha$ on the whole contour, integration
running over $\varphi'\,$, say, from $-\pi$ to $+\pi\,$. In the
$z$-plane, such a contour is an ellipse
\begin{equation}
\left(\frac{\mathrm{Re}\,z}{\cosh\alpha}\right)^2 +
\left(\frac{\mathrm{Im}\,z}{\sinh\alpha}\right)^2=1\,,
\label{ellips}
\end{equation}
having semiaxes $\cosh\alpha,\,\,\sinh\alpha$ and foci at $z=-1,\,
\,z=+1\,$; at the contour $$dz'=i\,\sinh(\alpha+i\,
\varphi')\cdot d\varphi'\,.$$

Now we can use integral (\ref{cint}) to investigate behavior of
$a_l$ at large values of $l\,$. The Legendre function $Q_l$
may be written in the form~\cite{HTF}
\begin{equation}
 Q_l(\cosh\beta)=\sqrt{\pi}\cdot\frac{\Gamma\left(l+1\right)}
{\Gamma\left(l+\frac32\right)}\cdot\frac{e^{-\beta(l+1)}}
{\sqrt{1-e^{-2\beta}}}\cdot{}_2F_1\left(\frac12,\,
\frac12;\,l+\frac32;\,\frac1{1-e^{2\beta}}\right)\,,
~~~~\mathrm{Re}\,\beta>0\,,
\label{ql}
\end{equation}
which provides the asymptotic expression
\begin{equation}
\sinh(\alpha+i\,\varphi')\cdot Q_l(z')\,|_{l\to+\infty}=
e^{-\left(\alpha+i\,\varphi'\right)\left(l+1/2\right)}
\cdot\sqrt{\frac{\pi}{2l}\,\sinh(\alpha+i\,\varphi')}\cdot
\left[1+{\cal{O}}\,\left(l^{-1}\right)\right]\,.
\label{asymq}
\end{equation}
Since $\sinh\alpha<|\sinh(\alpha+i\,\varphi')|<\cosh\alpha\,,$ for large real $l$ there is the upper boundary
\begin{equation}
|\sinh(\alpha+i\,\varphi')\,Q_l(z')\,|_{\,l\to+\infty}
<e^{-\alpha(l+1/2)}\,\sqrt{\frac{\pi}{2l}\,\cosh\alpha}\,\,,
\label{boundq}
\end{equation}
independent of $\varphi'\,.$ Therefore,
\begin{equation}
|a_l|<\frac{q_s}{\sqrt{s}}\cdot\,B_\alpha(s)\cdot
\sqrt{e^{-\alpha}\,\cosh\alpha}
\cdot\frac{e^{-\alpha\,l}}{\sqrt{\,l}}\,,
\label{boundam}
\end{equation}
where
\begin{equation}
B_\alpha(s)=\sqrt{\frac{\pi}{8}}\,
\int_{-\pi}^{+\pi} |A(s, z')|\,d\varphi'\,,
~~~~~z'=\cosh(\alpha+i\,\varphi')\,.
\label{bint}
\end{equation}
Evidently, the upper boundary (\ref{boundam}) decreases with
increasing $l\,$. The larger $\alpha$ is, the faster is the
decrease of the boundary, which means the stricter limitation
for $|a_l|$ at large $l\,$. If the amplitude $A(s,z)$ has a
singularity nearest to the physical region at
$$z=z_0\equiv\cosh(\alpha_0+i\,\varphi_0)\,,$$
then the contour in Eq.(\ref{cint}) may be blown up until
it touches this nearest singularity. If the singularity is
integrable and $B_\alpha$ stays finite in this limit, we can
drag $\alpha$ to $\alpha_0$ (the value of $\varphi_0$ may
influence only the limiting value of $B_\alpha\,$). For the
amplitude $A(s,z)\,,$ the ellipse with $\alpha=\alpha_0$ is
just the Lehmann ellipse of analyticity in $z$~\cite{lehm}.

Let us consider, in more detail, possible singularities in
the $z$-plane. Unitarity predicts that an amplitude has
singularities in both $s$-, and $t$- or $u$-channels.
Those may be poles, corresponding to one-particle states,
or branch points, corresponding to two-particle or
multiparticle thresholds. Any of such singularities has a
position described by a definite value of the corresponding
Mandelstam invariant, say, $t_0$, independent of other
Mandelstam invariants. In addition, there can be anomalous
singularities (Landau singularities)~\cite{landau}, whose
positions, say, in the $t$-plane, depend on the $s$-value.
However, each leading anomalous singularity in reaction
(\ref{react}) is related to some threshold, and at $s\to+\infty$
it tends toward the corresponding threshold point. Since we
are interested here just in large positive $s$, we will neglect
possibility of $s$-dependence for all $t$- or $u$-singularities
meaningful in our present problem.

According to Eq.(\ref{angle}), all $t$- and/or $u$-channel
singularities reveal themselves in the $s$-channel as
singularities in the $z$-plane. One-particle and threshold
singularities, related to stable particles,
have real non-negative values of $t$ or $u$. This means
that the $t$-channel ($u$-channel) generates real
$z$-singularities at $z>+1$ ($z<-1$). Unstable particles
generate complex singularities, but they are not leading
(nearest) ones, being positioned at secondary Riemann sheets
(we assume initial and final particles in reaction
(\ref{react}) to be stable). Anomalous singularities also
can be complex, but the nearest ones are real. Thus, in
conventional opinion, the nearest $z$-singularity has either
$\varphi_0=0\,$, for $t$-channel, or $\varphi_0=\pm\pi\,$,
for $u$-channel. As was explained, the $\varphi_0$-value
may influence the boundary (\ref{boundam}) only through
the coefficient $B$. In what follows, we assume, for
simplicity, that the nearest singularity is related to the
$t$-channel; it has $\varphi_0=0$ and positioned at fixed
point $t=t_0\ge0\,.$

If there exist massless particles, as in quantum electrodynamics
(QED), then some amplitudes may have a pole at $t_0=0$,
\textit{i.e.}, at the edge of the physical region, at $z=+1$.
Then both the corresponding forward amplitude and the total
cross section are infinite at any value of the $s$-channel
energy, their boundaries being meaningless. Multiphoton
exchanges are related to thresholds, also at $t_0=0$. Such
singularities are also at the edge of the physical region, but
they are integrable and do not provide infinities of the
forward amplitudes and/or total cross sections. Applicability
of high-energy boundaries for QED amplitudes without one-photon
exchanges needs special investigation.

If there are no massless particles, then $t_0>0$, and $z_0=
\cosh\alpha_0=1+(t_0/2q_s^2)>1\,.$ There is a finite
interval of $\alpha\,,$ from 0 to $$\alpha_0=\ln\left(z_0+
\sqrt{z_0^2-1}\,\right)\,,$$ where Eq.(\ref{cint}) is
applicable and the boundary (\ref{boundam}) is operative.
The limit $\alpha\to\alpha_0$ may be reached if the effective
value of $B_\alpha$ stays finite. Let us consider this problem
in some more detail. If $t_0$ corresponds to a pole, one can
separate the pole contribution and continue to blow up the
remaining contour (\ref{cint}) further, till the next
singularity. The pole term provides then the inequality
(\ref{boundam}) with $\alpha=\alpha_0$ and the coefficient $B$
expressed through the pole residue (below we will explicitly
consider this case). Of course, contribution of the continued
contour decreases with $l$ faster than the pole one. For the
nonpole leading singularity, both threshold and anomalous
leading singularities are integrable, and the corresponding
expression (\ref{bint}) is finite even at $\alpha=\alpha_0\,,$
when the singularity lies just at the integration contour.
Therefore, in all practical cases, one can use the boundary
(\ref{boundam}) with $\alpha=\alpha_0$ and some finite
coefficient $B_0(s)\,.$ At high energies $\alpha_0 \approx
\sqrt{t_0}/q_s\approx2\sqrt{t_0/s}\,.$ The factor
$$\sqrt{e^{-\alpha_0}\,\cosh\alpha_0}=
\sqrt{1-e^{-\alpha_0}\,\sinh\alpha_0}$$
is always lower than unity, and tends toward unity at high
energies. Therefore, for our purpose here (for finding upper
boundary at high energies), we can change this factor by unity.

Thus, after all we have two upper boundaries for the partial
amplitudes:
\begin{equation}
|a_l|\le1~~~~ \textrm{and}~~~~|a_l|<\frac{q_s}{\sqrt{s}}\,
\,B_0(s)\,\,\frac{e^{-\alpha_0\,l}}{\sqrt{\,l}}\,\,.
\label{pw2bound}
\end{equation}
The first inequality is true for any $l$, while the second one
is applicable only at sufficiently large values of $l\,$. It is
interesting to compare these inequalities for partial amplitudes
with those of Froissart~\cite{frois}.

Of course, the first inequality is the same in both cases. But
the second one is slightly different. Froissart's Eq.(4) may be
rewritten as
$$|a_l|<\frac{q_s}{\sqrt{s}}\,\,B_{\mathrm{Fr}}(s)\,\,
\frac{e^{-\alpha_0\,(l-N)}}{\sqrt{\,l-N\,}}\,,$$
with integer positive $N$ equal to the number of subtractions.
(Of course, $l>N$; it seems, that $N$ appeared here because
Froissart worked with the infinite integration intervals, so
his dispersion integrals were subtracted; our integrals
(\ref{cint}) and (\ref{bint}) run over the final
$\varphi'$-interval and need no subtractions.) Our parameter
$\alpha_0$ is simply related with Froissart's parameter $x_0$:
$$\alpha_0=\log\left(x_0+\sqrt{x_0^2-1}\,\right)\,.$$
The quantity $B_{\mathrm{Fr}}(s)$ is constructed differently
than our $B_0(s)\,$, but it is also linearly related with the
amplitude in nonphysical configurations.

Note that Froissart's factor $\exp[-\alpha_0(l-N)]/\sqrt{l-N\,
}\,$ is somewhat larger than our $\exp[-\alpha_0\,l]/\sqrt{l}
\,$. Therefore, his Eq.(4) is somewhat weaker than our second
boundary (\ref{pw2bound}). Moreover, to simplify further
calculations, Froissart additionally changed the factor
$1/\sqrt{l-N\,}<1$ by just unity (see Ref.~\cite{frois},
lower left column on p.1055). Thus, for the $l$-dependence in
the second inequality (\ref{pw2bound}) he effectively used the
purely exponential factor $\exp(-\alpha_0 l)$, instead of
the smaller factor \mbox{$\exp(-\alpha_0 l)\cdot l^{-1/2}\,$.}
Martin~\cite{mart} also applied softened boundaries for
the Legendre functions.

In difference, the present approach allows us to use inequality
(\ref{boundq}) which is the strictest boundary for the
asymptotics (\ref{asymq}). In what follows, we retain the
resulting boundaries (\ref{pw2bound}) for the partial-wave
amplitudes as they are, without any further simplifications.

\section{Boundaries for amplitudes and cross sections}

Evidently, at very large $l\,,$ the latter of boundaries
(\ref{pw2bound}) is stricter than the former. Let us denote
$L$ to be the minimal value of $l$, for which the former
boundary is above the latter (we assume $L$ to be sufficiently
large, so that both inequalities (\ref{pw2bound}) are
applicable near $L$). Then
$$\frac{q_s}{\sqrt{s}}\,\,B_0(s)\,\,\frac{e^{-\alpha_0\,L}}
{\sqrt{\,L}}\,<1<\,\frac{q_s}{\sqrt{s}}\,\,B_0(s)\,\,
\frac{e^{-\alpha_0\,(L-1)}}{\sqrt{\,L-1}}\,,$$ or
\begin{equation}
1<\frac{\sqrt{s}}{q_s}\cdot\frac{e^{\alpha_0\,L}\,\sqrt{L}}
{B_0(s)}<e^{\alpha_0}\,\sqrt{\frac{L}{L-1}}\,.
\label{L}
\end{equation}
We see that generally $L$ depends on energy $s\,$. If $B_0(s)$
increases with energy, so does the corresponding value of $L\,$.
In such a case, the interval between the upper and lower
boundaries (\ref{L}) shrinks. Then, at high energies we
can write
\begin{equation}
e^{\alpha_0\,L}\,\sqrt{L}=\frac12\,B_0(s)\,,
\label{L1}
\end{equation}
keeping in mind that the correct value of $L$ is the nearest
integer number above the solution of equality (\ref{L1}).

Note that Froissart also introduced the interfacial number
$L$~\cite{frois}. His value $L_{\mathrm{Fr}}$ also increases
with $s\,$, but is different from ours. Due to the stronger
second boundary (\ref{pw2bound}), our inequalities (\ref{L})
provide a lower value of $L$ than Froissart's $L_{\mathrm{Fr}}$.
Martin's value of $L$ (denoted as $\bar{L}$)~\cite{mart} is
also larger than ours.

Now, to construct various bounds for the scattering amplitude
(\ref{pw-ser}), first of all we separate its series into two
parts, below and above $L$ (again, similar to
Froissart~\cite{frois}), and then estimate
\begin{equation}
|A(s,\cos\theta_s)|\le\frac{\sqrt{s}}{\pi q_s}\,\left[\,
\sum_{l=0}^{L-1}\,(2l+1)\,|a_l(s)|\cdot|P_l(\cos\theta_s)|+
\sum_{l=L}^{\infty}\,(2l+1)\,|a_l(s)|\cdot|P_l(\cos\theta_s)|
\,\right]\,.
\label{sumbound}
\end{equation}
To the partial amplitudes in each part, we apply the
corresponding boundary (\ref{pw2bound}).

\subsection{Forward (backward) amplitude}

For the forward (or backward) amplitude, with $|P_l(\pm1)|=
1\,,$ we obtain
\begin{equation}
|A(s,\pm1)|<\frac{\sqrt{s}}{\pi q_s}\,\cdot\sum_{l=0}^{L-1}\,
(2l+1)+\,\frac{B_0(s)}{\pi}\,\cdot\sum_{l=L}^{\infty}\,
e^{-\alpha_0\,l}\,(2l+1)\,\,l^{-\frac12}\,\,.
\label{fbound}
\end{equation}
The first term sums to $L^2\,\sqrt{s}/(\pi q_s)\,.$
The second term, with $l=L+l'$, can be rewritten as
$$\frac1{\pi}\,B_0(s)\,\frac{e^{-\alpha_0\,L}}{\sqrt{L}}
\cdot\sum_{l'=0}^{\infty}\,e^{-\alpha_0\,l'}\left[\,
2L\left(1+\frac{l'}{L}\right)^{\frac12}+\left(1+
\frac{l'}{L}\right)^{-\frac12}\right]\,, $$
which is, due to the left inequality (\ref{L}), smaller than
$$ \frac{\sqrt{s}}{\pi q_s}\,
\cdot\sum_{l'=0}^{\infty}\,e^{-\alpha_0\,l'}\left[\,
2L\left(1+\frac{l'}{L}\right)^{\frac12}+\left(1+
\frac{l'}{L}\right)^{-\frac12}\right]\,. $$
This sum converges only due to the decreasing exponential
factor. If we define $y=\alpha_0\,l',~Y=\alpha_0\,L\,$ then
at small $\alpha_0$ (\textit{i.e.}, at high energy) the sum
tends toward the integral (see Appendix)
\begin{equation}
I(s)=\frac1{\alpha_0} \int_0^\infty dy\,
e^{-y}\left[\,2L\left(1+\frac{y}{Y}\right)^{\frac12}+
\left(1+\frac{y}{Y}\right)^{-\frac12}\right]\,.
\label{fint}
\end{equation}
Thus, our boundary for the forward (backward) amplitude
takes the form
\begin{equation}
|A(s,\pm1)|<\frac{\sqrt{s}}{\pi q_s}\,\left[\,L^2+
I(s)\,\right]\,.
\label{fbound1}
\end{equation}
Its high-energy behavior directly depends on properties
of $L$ (and $Y=\alpha_0 L\,$).

Let us consider various possibilities. A finite limit of $L$
at $s\to\infty$ would mean that $B_0(s)$ has also a finite
limiting value. Such a case would lead to decreasing total
cross section and is not interesting here.

If $L$ increases with $s$, but $Y$ stays finite (or even
decreases), then $B_0(s)$ also grows, but not faster than
$(1/\alpha_0)^{1/2}\sim s^{1/4}$. The corresponding total
cross section cannot infinitely grow with energy. It tends
to constant (or may even slowly decrease in asymptotics).

In connection with the Froissart theorem, the most interesting
is the case when the total cross section does increase with
energy, without any finite limit. Then both $L$ and $Y$ should
grow. Integral (\ref{fint}) can then be approximately calculated
as $$I(s)=\frac1{\alpha_0^2}\,\left[2Y+1+{\cal{O}}(1/Y)\right],$$
and boundary (\ref{fbound1}) at high energies takes the simple
form
\begin{equation}
|A(s,\pm1)|<\frac2{\pi}\,\left(\frac{Y+1}{\alpha_0}\,
\right)^2\approx\frac{2q_s^2}{\pi t_0}\,\,(Y+1)^2\approx
\frac{s}{2\pi t_0}\,\,(Y+1)^2\,.
\label{fbound2}
\end{equation}
Therefore, at forward or backward angles, the modulus of the
amplitude behaves the most like $s\,Y^2$, when $s\to+\infty\,$.
If the considered amplitude corresponds to elastic scattering,
then one can use the optical theorem to derive that the total
cross section behaves the most like $Y^2$, as $s$ goes to
infinity.

\subsection{Fixed-angle amplitude}

For the fixed-angle configuration we will consider only the case
of an infinitely increasing total cross section, corresponding
to growing values of $L$ and $Y$. At nonforward (nonbackward)
angles we also begin with the inequality (\ref{sumbound}). Since
$L$ is growing with $s$, we can fix some finite number $l_0$ and
subdivide the first term, again into two parts, with $0\le
l<l_0$ and with $l_0\le l<L$. We choose the value $l_0$ so that
$P_l(\cos{\theta)}$ for $l\ge l_0$ can be, with good accuracy,
presented in its large-$l$ asymptotic form~\cite{HTF}. This
form corresponds to combining Eqs.(\ref{jump}) and
(\ref{asymq}); for $\epsilon<\theta<\pi-\epsilon$ it provides
the estimate
\begin{equation}
|\,P_l(\cos{\theta})\,|_{\,l\to+\infty}=\,\sqrt{\frac2{\pi l\,
\sin{\theta}}}\,\cdot\,\left|\,\left\{\,\cos\left[\left(l+
\frac12\right)\,\theta-\frac{\pi}4\,\right]+{\cal{O}}\left(l^{-1}
\right)\right\}\right|\,
<\sqrt{\frac2{\pi l\,\sin{\theta}}}\,\,.
\label{pasymp}
\end{equation}

In Eq.(\ref{sumbound}), contributions with $l<l_0$
do not grow at $s\to+\infty$. Thus, the amplitude increasing
at high energy should be related to (and bounded by) two sums:
\begin{equation}
|A(s,\cos\theta_s)|<\frac{\sqrt{s}}{\pi q_s}\cdot\sqrt{
\frac2{\pi\sin{\theta_s}}}\,\,\sum_{l=l_0}^{L-1}\,\,
(2\,l^{\frac12}+l^{-\frac12})+\frac{B_0(s)}{\pi}\cdot
\sqrt{\frac2{\pi\sin{\theta_s}}}\,\,\sum_{l=L}^{\infty}\,
e^{-\alpha_0\,l}\,(2+l^{-1})\,.
\label{nfbound}
\end{equation}
Just as for the forward amplitude, one can use the left
inequality (\ref{L}) to rewrite
$$|A(s,\cos\theta_s)|<\frac{\sqrt{s}}{\pi q_s}\cdot
\sqrt{\frac2{\pi\,\sin{\theta_s}}}\,\,\left[\,
\sum_{l=l_0}^{L-1}\,\,(2\,l^{\frac12}+l^{-\frac12})+
\sum_{l'=0}^{\infty}\,e^{-\alpha_0\,l'}\left(2\,
L^{\frac12}+\frac{L^{-\frac12}}
{1+l'/L}\right)\,\right].$$
The most singular high-energy behavior of these sums
is determined by their first terms. At $s\to+\infty$ they
can be approximated as (again, see Appendix)
$$|A(s,\cos\theta_s)|<\frac{\sqrt{s}}{\pi q_s}\cdot
\sqrt{\frac2{\pi\,\sin{\theta_s}}}\,\,\left(\,
\sum_{l=l_0}^{L-1}\,\,2\,l^{\frac12}+2\,L^{\frac12}
\sum_{l'=0}^{\infty}\,e^{-\alpha_0\,l'}\,\right)$$
$$~~~~~~~~~~~\approx\frac{\sqrt{s}}{\pi q_s}\cdot
\sqrt{\frac2{\pi\,\sin{\theta_s}}}\,\,\left(\,
\frac2{\alpha_0^{\frac32}}\,\int_0^Y dy\,\sqrt{y}
+\frac{2\,\sqrt{L}}{1-e^{-\alpha_0}}\right)\,,$$
which takes the final form
\begin{equation}
|A(s,\cos\theta_s)|<\frac{2\sqrt{s}}{\pi q_s}\cdot
\sqrt{\frac2{\pi\,\sin{\theta_s}}}\left(\frac{Y}{\alpha_0}
\right)^{\frac32}\left(\frac23+\frac1{Y}\right)\,.
\label{nfbound1}
\end{equation}
Evidently, the main term in the fixed-angle bound comes from the
sum over $l<L$, just as for the forward (backward) amplitude.
Presence of cosine in the asymptotic expression (\ref{pasymp})
gives evidence for possibility of oscillating angular
distributions in elastic scattering with increasing total
cross-section. If so, boundary (\ref{nfbound1}) limits the
upper edges of those oscillations.

Singularities of this boundary at the ends of the physical angular
interval, at $\theta=0$ or $\pi$, are related, of course, to the
change in the possible energy behavior: $\sim (q_s\,Y)^{3/2}$
(or $\sim L^{3/2}$) inside the interval and $\sim (q_s\,Y)^{2}$
(or $\sim L^2$) at its ends.

To understand how this works, let us consider in more detail
our boundary at very small angles. Note that near $\theta=0$
(\textit{i.e.}, $\cos{\theta}$ near unity) both all $P_l(\cos{
\theta})$ and all their derivatives are positive, and we may
eliminate signs of modulus for the Legendre polynomials in the
right-hand side of inequality (\ref{sumbound}). Moreover, one
can differentiate it over $\cos{\theta_s}$. The adequate
expression for $P_l(z)$ near $z=1$ is~\cite{HTF}
\[
P_l(z)={}_2F_1\left(l+1,\,-l;\,1;\,\frac{1-z}2\right)\,.
\]
It shows that every additional differentiation of $P_l(z)$ over $z$
at $z=1$ provides an additional factor which is quadratic in $l$.

Derivation of boundaries, considered above, shows that most
efficient in the case of the growing cross section are values
$l\sim L$. Therefore, the $n$-th derivative of the boundary over
$\cos{\theta_s}$ at $\cos{\theta_s}=1$ grows with energy as
$(L^2)^{n+1}$. This means that the angular dependence of the
boundary reveals a narrow forward peak which rapidly shrinks
with energy growing. Formally, the same is true for the backward
scattering. We will return to this situation when considering
the fixed-$t$ configuration.

\subsection{Fixed-$t$ (or -$u$) amplitude}

Up to now, following to Froissart~\cite{frois}, we considered
high-energy boundaries for amplitudes in two regimes: either
forward (backward) amplitudes, related in elastic cases to
the total cross sections; or amplitudes of two-particle
processes at a fixed angle. Here we consider another interesting
regime, not discussed by Froissart or any other author. It is
the case of fixed-momentum transfer, $t$ or $u$. We begin,
again, with two-term Eq.(\ref{sumbound}).

For definiteness, let us take at first a fixed value of $t$ and denote the
corresponding amplitude as $A(s,t)$. At high energy, according
to Eq.(\ref{angle}), the corresponding $\cos{\theta_s}\to1$,
\textit{i.e.}, \mbox{$\theta_s\to0\,$.} Then, one can
approximately express the Legendre polynomials with sufficiently
large $l\ge l_0$ through the Bessel functions~\cite{HTF}
\begin{equation}
P_l(\cos{\theta_s})=J_0(\xi)+{\cal O}\,(\theta_s^2)\,,
\label{pbes}
\end{equation}
with $\xi=(l+1/2)\,\theta_s\,$. At high energy, $\xi\approx
y\,\sqrt{(-t)/t_0}\,$, where again $y=\alpha_0\,l\,$.

Now, in analogy with previous subsections, we obtain the estimate
$$ |A(s,t)|<\frac{\sqrt{s}}{\pi q_s}\,\cdot\sum_{l=l_0}^{L-1}\,
(2l+1)\,|J_0(\xi)| +\,\frac{B_0(s)}{\pi}\,\cdot
\sum_{l=L}^{\infty}\,e^{-\alpha_0\,l}\,(2l+1)\,\,
l^{-\frac12}\,|J_0(\xi)|\,,$$
which can, after using the left inequality (\ref{L}) and the
fact that essential values of $l-L$ at high energy are much
less than $L$, be rewritten as
$$ |A(s,t)|<\frac{\sqrt{s}}{\pi q_s}\,\left[\sum_{l=l_0}^{L-1}\,
(2l+1)\cdot|J_0(l\,\theta_s)| +\sum_{l'=0}^{\infty}\,e^{-\alpha_0\,l'}\,
(2L+l'+1)\cdot|J_0(L\,\theta_s+l'\theta_s)|\right]\,$$
(compare to inequality (\ref{fbound}) and its transforms).
High-energy asymptotics of the right-hand side can be, again,
expressed through integrals
$$ |A(s,t)|<\frac{\sqrt{s}}{\pi q_s}\cdot\frac2{\alpha_0^2}\cdot
\left[\int_0^{Y} dy\cdot y\,\left|J_0\left(y\,\sqrt{\frac{-t}
{t_0}}\,\right)\right| + \int_0^{\infty}dy'\cdot e^{-y'}\,
\left(Y+\frac{y'}{2Y}\right)\left|J_0\left(\frac{Y+y'}{\sqrt{t_0}}
\sqrt{-t}\right)\right|\right]\,.$$

As in other cases considered untill now, the second integral here
is less singular at high energies and can be neglected. Then,
our final high-energy estimate for the fixed-$t$ amplitude is
\begin{equation}
|A(s,t)|<\frac2{\pi}\cdot\frac{Y^2}{\alpha_0^2}
\cdot\int_0^1 dx\,\left|J_0\left(Y\,\sqrt{\frac{(-t)\,x}
{t_0}}\,\right)\right|\,,
\label{tbound}
\end{equation}
with $x=(y/Y)^2\,$ (the limit $\sqrt{s}/q_s\to2$ at $s\to\infty
\,$ is also used here). At $t=0$ this inequality coincides with
the estimate (\ref{fbound2}) for the forward amplitude. At
non-zero finite values of $t$ the argument of the Bessel
function infinitely increases with energy, and we can use the
asymptotic expression~\cite{JEL}
\begin{equation}
|\,J_0(\xi)\,|_{\xi\to+\infty}=\sqrt{\frac2{\pi\xi}}\,\cdot
\left|\,\left[\,\cos\left(\,\xi-\frac{\pi}4\,\right)
+{\cal{O}}\left(\xi^{-1}\right)\,\right]\right|
<\sqrt{\frac2{\pi\xi}}
\label{besasym}
\end{equation}
(compare to Eqs.(\ref{pasymp}) and (\ref{pbes}) )\,.
Then, finally, we obtain
\begin{equation}
|A(s,t)|<\frac4{3\,\alpha_0^2}\left(\frac{2Y}{\pi}\right)^{
\frac3{\,2}}
\cdot\left(\frac{t_0}{-t}\right)^{\frac1{\,4}}\,.
\label{tbound1}
\end{equation}
Energy behavior of this boundary ($\sim Y^{3/2}/\alpha_0^2$)
is intermediate between the forward boundary \mbox{($\sim
Y^2/\alpha_0^2$)} and the fixed-angle one ($\sim
Y^{3/2}/\alpha_0^{3/2}$). It shows also a rather slow decrease
with increasing momentum transfer, $\sim (-t)^{-1/4}$.

Again, similar to the fixed-angle case, presence of cosine in
Eq.(\ref{besasym}) may provide evidence for a possible
oscillating $t$-distribution. Then the boundary (\ref{tbound1})
limits the upper edges of those oscillations.

Similar to the fixed-angle case at $\theta_s\to0\,$, the
fixed-$t$ bound is singular at $(-t)\to0\,$. This singularity,
again, is spurious, related to the change of the energy
behavior. The boundary (\ref{tbound}) for the amplitude in the
small $|t|$-region increases with energy very differently at
$t=0\,$ ($\sim Y^2/\alpha_0^2$) and at small finite value of
$|t|\,$ ($\sim Y^{3/2}/\alpha_0^2$). This means that the
amplitude boundary (\ref{tbound}) at small $|t|$ reveals a
narrow peak which shrinks when energy grows.

To understand the structure of this peak, let us consider
in more detail relation
$$|A(s.t)|< A^{\mathrm{(max)}}(s,t)\,.$$
 Near $t=0$ the $A^{\mathrm{(max)}}(s,t)$
corresponds to the right-hand side (\ref{tbound}). At
sufficiently small $|t|$ and fixed $s$ (and $Y$ as well)
the Bessel function in expression (\ref{tbound}) is positive,
and the signs of modulus may be omitted. For small
arguments~\cite{JEL}
$$J_0(z)\approx 1-\frac{z^2}4\,,$$
and we obtain
\begin{equation}
\frac{d~}{dt}A^{\mathrm{(max)}}(s,t)|_{\,t=0}=
\frac2{\pi}\cdot\frac{Y^2}{\alpha_0^2}\cdot\frac{Y^2}
{8t_0}=A^{\mathrm{(max)}}(s,0)\cdot\frac{Y^2}{8t_0}\,.
\label{deriv}
\end{equation}

When considering a differential cross section, it is familiar
to parameterize its near-forward \mbox{$t$-dependence} as
$\exp(bt)$ (recall that $t<0$ in the physical region and that we
do not account for spins). Since $d\,\sigma(s,t)/d\,t\,\propto\,
|A(s,t)|^2$, we can express the slope of the forward peak as
\begin{equation}
b=\frac{d~~}{d\,t}\log\left[\frac{d\,\sigma(s,t)}{dt}
\right]_{t=0}=2\,\frac{d~~}{d\,t}\log\left|A(s,t)
\right|_{\,t=0}\,.
\label{slope}
\end{equation}
In analogy, we can define the slope related to the boundary
$A^{\mathrm{(max)}}(s,t)$ as
\begin{equation}
b^{\mathrm{(max)}}=2\,\frac{d~~}{d\,t}\log\left[
A^{\mathrm{(max)}}(s,t)\right]_{\,t=0}\,.
\label{slopem}
\end{equation}
Of course, $b^{\mathrm{(max)}}$ does not necessarily provide
a bound for $b$, though $A^{\mathrm{(max)}}$ is the bound
for $|A|$. The reason is evident: differentiation may violate
inequalities. Now, Eq.(\ref{deriv}) shows that at high
energies $b^{\mathrm{(max)}}=Y^2/(4\,t_0)$. It is
interesting to note that the slope $b^{\mathrm{(max)}}$
has the high-energy behavior $\sim Y^2$, exactly the same
as $\sigma_{\mathrm{tot}}^{\mathrm{(max)}}$, the boundary for
$\sigma_{\mathrm{tot}}\,$. Therefore,
$$\left[\,\frac{\,b^{\mathrm{(max)}}}{\,
\sigma_{\mathrm{tot}}^{\mathrm{(max)}}}\,\right]_{\,s\to+\infty}=
\mathrm{const}\,.$$
There is, however, an essential difference between the two
quantities: the physical total cross section
$\sigma_{\mathrm{tot}}$ is always bounded by
$\sigma_{\mathrm{tot}}^{\mathrm{(max)}}$, while the
physical slope $b$ may be either larger or smaller than
$b^{\mathrm{(max)}}\,$. But if $\sigma_{\mathrm{tot}}$ is
saturated and indeed increases $\sim Y^2$, then the diffraction
peak slope $b$ should be also saturated and increase with
energy not slower than $b^{\mathrm{(max)}}\,$. Thus, in the
saturated regime $b\ge b^{\mathrm{(max)}}\,$. Otherwise, the
amplitude boundary (\ref{tbound}) at fixed $t<0$ might become
violated when energy grows.

The high-energy scattering at fixed $t$ corresponds to angles
near $\vartheta_s=0\,$, \textit{i.e.}, to forward scattering.
The case of backward scattering, for angles near $\vartheta_s=
\pi\,$, may be considered in a similar way, with change $t\to
u\,$.

\subsection{Amplitude inside the Lehmann ellipse}

Our approach allows us to discuss one more case, asymptotics
of the amplitude outside the physical region, but inside the
Lehmann ellipse. Though this case is not of direct physical
interest, it may have theoretical interest. Here we again apply
Eq.(\ref{sumbound}), but instead of $P_l(\cos\vartheta_s)$ we
use $P_l(z)$ with $z=\cosh(\alpha+i\varphi)$; by our convention,
$\alpha>0$.

To find the large-$l$ asymptotics of $P_l(z)$ with $z$ outside
the physical region, we can use the relation between Legendre
functions of the 1st and 2nd kinds~\cite{HTF}:
\begin{equation}
Q_l(\cosh\beta)-Q_{-l-1}(\cosh\beta)=\pi\,\frac{\cos(\pi l)}
{\sin(\pi l)}\cdot P_l(\cosh\beta)\,.
\label{leg1-2}
\end{equation}
Substituting the corresponding expressions (\ref{ql}) and
tending the value of $l$ to a positive integer number, we
obtain relation
\begin{equation}
P_l(\cosh\beta)=\frac{\Gamma(l+\frac12)}
{\sqrt{\pi}\,\Gamma(l+1)}\cdot\frac{e^{\beta l}}
{\sqrt{1-e^{-2\beta}}}\cdot{}_2F_1\left(\frac12,\,
\frac12;\,-l+\frac12;\,\frac1{1-
e^{2\beta}}\right)\,,~~~~\mathrm{Re}\,\beta>0\,.
\label{asymp}
\end{equation}
Of course, this relation is correct only at positive integer
values of $l$, but only such values appear in the sums
(\ref{sumbound}). For large $l$, it provides the asymptotic form
$$P_l(\cosh\beta)\,|\,_{l\to+\infty}=\frac1{\sqrt{2\pi\,l}}\cdot
\frac{e^{\beta(l+\frac1{\,2})}}{\sqrt{\sinh\beta}}\cdot[1+
{\cal{O}}(l^{-1})]\,,~~~~\mathrm{Re}\,\beta>0\,,$$
which, for $z=\cosh(\alpha+i\varphi)$, gives
\begin{equation}
|P_l(z)|_{\,l\to+\infty}=\frac1{\sqrt{2\pi\,l}}\cdot
\frac{e^{\alpha(l+\frac1{\,2})}}{(\sinh^2\alpha+
\sin^2\varphi)^{\frac1{\,4}}}\cdot[1+{\cal{O}}(l^{-1})]\,,
~~~~\alpha>0\,.
\label{boundp}
\end{equation}

If we use this expression in the sums
\begin{equation}
|A(s,z)|\le\frac{\sqrt{s}}{\pi q_s}\,\left[\,
\sum_{l=0}^{L-1}\,(2l+1)\,|a_l(s)|\cdot|P_l(z)|+
\sum_{l=L}^{\infty}\,(2l+1)\,|a_l(s)|\cdot|P_l(z)|
\,\right]
\label{sumboundl}
\end{equation}
and apply, as earlier, bounds (\ref{pw2bound}), we see
that the second sum converges only at $\alpha<\alpha_0$ (recall
that $\alpha=\alpha_0$ is the singular point where the series
should diverge). Since $\alpha_0\to0$ at $s\to+\infty\,$, our
approach does not allow to investigate high-$s$ behavior
of $A(s,z)$ at any fixed $z$ outside the physical region.
However, we are able to consider, \textit{e.g.}, the case of
a fixed nonphysical value for the momentum transfer $t\,$ (or
$u$).

The nonphysical interior of the $z$-plane Lehmann ellipse
corresponds to $0\le\alpha\le\alpha_0$ and is described by
inequality
\begin{equation}
\left(\frac{\mathrm{Re}\,z}{\cosh\alpha_0}\right)^2 +
\left(\frac{\mathrm{Im}\,z}{\sinh\alpha_0}\right)^2\le1\,
\label{ellips0}
\end{equation}
(compare to Eq.(\ref{ellips})). Since $z=1+t/(2q_s^2)\,$
and $\alpha_0\approx\sqrt{t_0}/q_s$ at high energies, we can
rewrite condition (\ref{ellips0}) for $s\to+\infty$ as
\begin{equation}
\frac{\mathrm{Re}\,t}{t_0}+
\left(\frac{\mathrm{Im}\,t}{2\,t_0}\right)^2\le1\,.
\label{ellipst}
\end{equation}
The case of equality here corresponds to the limiting form of
the Lehmann ellipse. In the complex $t$-plane, it is the
parabola, symmetrical with respect to the real $t$-axis and
directed to the left of $t=t_0\,$.

For any point $t$ inside this parabola one can define
\begin{equation}
t_r=(|t|+\mathrm{Re}\,t)/2\,.
\label{tr}
\end{equation}
Then,
$$\frac{\mathrm{Re}\,t}{t_r}+
\left(\frac{\mathrm{Im}\,t}{2\,t_r}\right)^2=1$$
(compare to Eq.(\ref{ellipst})). Inside the parabola,
$0\le t_r\le t_0\,$. Further, we can use the familiar relation
$z=1+t/(2q_s^2)\,$. At high (but not infinite) $s$-values and
fixed $t\,$, the parametrization $z=\cosh(\alpha+i\,\varphi)$
provides $\alpha\approx\sqrt{t_r}/q_s\approx\alpha_0\cdot
\sqrt{t_r/t_0}\,$, $\,\,|\varphi|\approx\sqrt{(t_r-
\mathrm{Re}\,t)\,}/q_s\,\approx\alpha_0\cdot\sqrt{(t_r-
\mathrm{Re}\,t)/t_0}\,$, and $\,\sinh(\alpha+i\,\varphi)
\approx\sqrt{t\,}/q_s\,$. Expression (\ref{boundp}) takes the form
\begin{equation}
|P_l(z)|_{\,l,\,s\to+\infty}\approx\frac1{\sqrt{2\pi\,l\,
\alpha_0}}\cdot e^{\alpha_0\,l \sqrt{t_r/t_0}}\cdot\left(
\frac{t_0}{\,|t|}\right)^{\frac1{\,4}}\,.
\label{boundp1}
\end{equation}
Note that in the physical region, at real $t\le0\,$, $\alpha=0$
and $|\varphi|=\theta_s\,$, as should be.

Now we can return to the inequality (\ref{sumboundl}) and
continue construction of the boundary for the amplitude
$A(s,t)$. As before, contributions of finite-$l$ terms are
inessential for the case of increasing $\sigma_{tot}$
(\textit{i.e.}, for increasing $L$), and we will run the
summation from some $l=l_0$, which is fixed, but sufficiently
large to admit application of the asymptotics (\ref{boundp}),
(\ref{boundp1}). Then
$$ |A(s,t)|\le\frac2{\pi}\left(\frac{\,t_0}{\,|t|}
\right)^{\frac14}\,\left[\,\sum_{l=l_0}^{L-1}\,
\frac{2l+1}{\sqrt{2\pi\,l\,\alpha_0}} \,e^{\alpha_0\,l
\sqrt{t_r/t_0}}+\sum_{l=L}^{\infty}\, \frac{2l+1}{\sqrt{2\pi\,
l\,\alpha_0}} \,B_0(s)\,\,\frac{e^{-\alpha_0\,l}}{2\sqrt{\,l}}
\,\,e^{\alpha_0\,l\sqrt{t_r/t_0}}\,\right]\,.$$
Using for $B_0(s)$ the left inequality (\ref{L}) (or,
equivalently, equality (\ref{L1})), changing the sums by
integrals (as described in Appendix and used in the previous
subsections), and discarding inessential contributions,
we obtain
$$|A(s,t)|<\left(\frac2{\pi}\right)^{\frac3{\,2}}\left(\frac{t_0}
{|t|}\right)^{\frac1{\,4}}\,\left[\frac1{\alpha_0^2}\,\int_{0}^{Y}
dy\,\sqrt{y}\,\,e^{\,y\sqrt{t_r/t_0}}+\frac{\sqrt{Y}\,
e^{Y\sqrt{t_r/t_0}}}{\alpha_0^2}\int_{0}^{\infty}dy\,\,
e^{-y(1-\sqrt{t_r/t_0}\,)}\,\right]\,.$$
The second integral here diverges at $t_r\to t_0$ and, thus,
restricts the region of applicability for our bound as
$t_r<t_0\,$. Now, integration of the second term and
transformation of the integral in the first term provides
the boundary
\begin{equation}
|A(s,t)|<\frac1{\alpha_0^2}\,\left(\frac{2Y}{\pi}
\right)^{\frac3{\,2}}\left(\frac{t_0}{|t|}\right)^{\frac1{\,4}}\,
\left[\frac23\int_{0}^{1}dx\,\,e^{\,x^{2/3}\,Y\sqrt{t_r/t_0}}
+\frac{\,e^{Y\sqrt{t_r/t_0}}}{Y(1-\sqrt{t_r/t_0\,}\,)}
\,\right]\,,
\label{tbound2}
\end{equation}
where $x=(y/Y)^{3/2}$.

It is interesting to compare this boundary with the similar
fixed-$t$ boundary (\ref{tbound1}) for real negative $t$-values
of the physical region. As was explained, the physical region
may be reached by the limit $t_r\to0$ at $|t|\neq0\,$. When we
formally apply this limit to the boundary (\ref{tbound2}),
the second term becomes a parametrically small ($\sim 1/Y$)
correction to the first one, and we can neglect it. Then the
boundary (\ref{tbound2}) takes the same functional structure
as the boundary (\ref{tbound1}), but being twice as small. The
difference can be traced to the different large-$l$ asymptotics
of the Legendre polynomials $P_l(z)$ inside the physical
interval ($z=x,\, -1<x<+1\,$) and outside it. The cosine,
present in the asymptotic expression (\ref{pasymp}) (and in
the related expression (\ref{besasym})), is the combination of
two exponentials, both of which should be taken into account.
Only one of those exponentials is essential in the similar
asymptotic expression (\ref{boundp}), for $z$ outside the
physical region. Therefore, transition between asymptotics
inside and outside the physical region may be non-continuous
(the Stokes phenomenon). Note, however, that the modulus of
the cosine reveals oscillations, which frequency increases
with energy. Their averaging provides the factor $1/2\,$ and
could make the asymptotics to be continuous between physical
and nonphysical regions.

To find explicit high-energy behavior of the boundary
(\ref{tbound2}), we need to calculate its integral. Using
decomposition for the exponential, we obtain~\cite{HTF}
$$\int_{0}^{1}dx\,\,e^{\,x^{2/3}\,Z} =\sum_{n=0}^{\infty}
\,\frac{Z^n}{n!}\cdot\frac1{\frac23\,n+1}=\Phi(\frac32,\,
\frac52;\,Z)\,,~~~~\Phi(\frac32,\,\frac52;\,Z){}_{\,Z\to+
\infty}=\frac32\cdot\frac{e^Z}{Z}\,[1+{\cal{O}}(Z^{-1})]\,,$$
where $Z=Y\sqrt{t_r/t_0}\,$. Finally, at fixed $t$ with
$0<t_r<t_0$, we obtain
\begin{equation}
|A(s,t)|_{\,s\to+\infty}<\left(\frac2{\pi}\right)^{\frac3{\,2}}
\left(\frac{t_0}{\,|t|\,}\right)^{\frac1{\,4}}\cdot
\frac{\sqrt{Y}}{\alpha_0^2}~e^{\,Y\sqrt{t_r/t_0}}\cdot
\left(\frac1{\sqrt{t_r/t_0}}+\frac1{1-\sqrt{t_r/t_0}}
\right)\,.
\label{tbound3}
\end{equation}
Thus, after all, the high-energy asymptotics is not continuous
between physical and nonphysical regions. Presence of
singularities in the boundary (\ref{tbound3}) at $t_r=0$ and
$t_r=t_0$, similar to previous cases, is related to the change
of the asymptotic behavior.

It is interesting to note  difference between bounds for the
nonphysical interior of the Lehmann ellipse and all the cases
considered before. The second sum (contributions of partial
waves with $l\ge L$; in the brackets of Eq.(\ref{tbound3}) they
provide the second term) has asymptotically the same functional
behavior with respect to energy-dependent parameters $Y$ and
$\alpha_0$, as the first sum (coming from waves with $l<L$;
in the brackets of Eq.(\ref{tbound3}) see the first term).
Numerically, the second term becomes even larger than the first
one, if $t_r$ is close to the singularity  point $t_0\,$. In the
physical region, contribution of waves with $l\ge L$ was
parametrically smaller than contribution of waves with $l<L$.

Evidently, the amplitude in the nonphysical region can grow
much faster than in the physical region, either inside it or
at the edge. This is directly related to difference in the
high-$l$ asymptotics of the Legendre polynomials $P_l(z)\,$.
In the whole physical region, they all do not exceed unity.
Moreover, as a function of $l$ inside the physical region,
they oscillate at physical arguments and slowly decrease with
growing $l\,$. On the other hand, outside the physical region,
$P_l(z)$ exponentially increases with growing $l\,$. This
is just the origin of the exponential factor in the bound
(\ref{tbound3}).

In this subsection, we have studied the case of fixed $t$
inside the Lehmann ellipse. In terms of the $z$-plane, this
corresponds to $z\to+1\,$. The case of fixed $u$-value, which
corresponds to $z\to-1\,$, may be considered in a similar
way. However, we can discuss also other points inside the
Lehmann ellipse. They may be characterized by
$$t-u=4q_s^2\, z$$
(recall that $t+u=-s+4m^2$). Then the condition (\ref{ellips0})
for the ellipse interior takes the form
\begin{equation}
\left[\frac{\mathrm{Re}\,(t-u)}{4q_s^2\,\cosh\alpha_0}\right]^2
+\left[\frac{\mathrm{Im}\,(t-u)}{4q_s^2\,\sinh\alpha_0}\right]^2
<1\,,
\label{ellipsut}
\end{equation}
which shows that Re$\,(t-u)$ and Im$\,(t-u)$ inside the ellipse
can grow with energy not faster than $\sim q_s^2$ and $\sim
q_s$ correspondingly. The fixed-$t$ (or -$u$) case is just a
particular case of such extreme possibility. Now, if we
parametrize, as before, $z=\cosh(\alpha+i\varphi)=\cosh\alpha
\cdot\cos\varphi+i\sinh\alpha\cdot\sin\varphi\,$, then
$$\mathrm{Re}\,(t-u)=4q_s^2\,\cosh\alpha\cdot\cos\varphi\,,~~
~~\mathrm{Im}\,(t-u)=4q_s^2\,\sinh\alpha\cdot\sin\varphi\,.$$

Since $0<\alpha<\alpha_0$ inside the Lehmann ellipse, $\alpha$
should decrease at $s\to+\infty$ as $\,\sim q_s^{-1}$ or
faster, while $\varphi$ may stay fixed. In this limit $z\to
\cos\varphi\,$, and $|\varphi|$ may be confronted with
$\theta_s\,$. If we construct a high-energy boundary for the
fixed-$\varphi$ amplitude, starting from $\alpha>0$ and using
the asymptotic expression (\ref{boundp}), the exponential
factor will contain the parameter $\alpha\,L=(\alpha/\alpha_0)
\,Y\,$. For the case of $\alpha$ decreasing faster than
$q_s^{-1}$, the boundary appears to have the same functional
structure as the fixed-angle one (\ref{nfbound1}), but being
twice as small (compare with the relation between the two
fixed-$t$ boundaries, physical boundary (\ref{tbound1}) and
nonphysical one (\ref{tbound2}) at $t_r\to0$).

For $\alpha$ decreasing as $q_s^{-1}$, it is reasonable again
to define $t_r$ by the relation
\begin{equation}
1+\frac{t_r}{2\,q_s^2}=\cosh\alpha=
\frac1{4\,q_s^2}\,(\,|t|+|u|\,)\,.
\label{tr1}
\end{equation}
The latter equality results from Eq.(\ref{ellips}) and
relations between $t,u$ and $z\,$. The value $t_r$ defined in
such a way is the $t$-value corresponding to the real positive
$z$-point being on the same $z$-plane ellipse as the given
point $t$ (or $u\,$). The earlier expression (\ref{tr}) for
$t_r$ arises as the high-energy limit of Eq.(\ref{tr1})
at fixed $t$. As in the nonphysical fixed-$t$ case, we have
$0<t_r<t_0$ and, in the high-energy limit,
$$\alpha\approx\frac{\sqrt{t_r}}{q_s}\,,~~~~
\frac{\alpha}{\alpha_0}\approx\sqrt{\frac{\,t_r}{\,t_0}}\,.$$

Now, to the case of fixed $\varphi$ and $t_r$, we can apply
the same procedure as described above for the fixed-$t$ case,
starting from Eqs.(\ref{boundp}) and (\ref{sumboundl}) to
construct the high-energy boundary. In this manner we obtain
$$|A(s,z)|<\left(\frac2{\pi\,\alpha_0}\right)^{\frac3{\,2}}
\left(\sin^2\varphi+\frac{t_r}{t_0}\alpha_0^2\right)^{-
\frac1{\,4}}~~~~~~~~~~~~~~~~~~~~~~~~~~~~~~~~~~~~$$
$$~~~~~\times\left[\int_{0}^{Y} dy\,\sqrt{y}\,\,
e^{\,y\sqrt{t_r/t_0}}+\sqrt{Y}\,e^{Y
\sqrt{t_r/t_0}}\,\int_{0}^{\infty}dy\,\,e^{-y(1-\sqrt{t_r/t_0}
\,)}\,\right]\,.$$
High-$Y$ asymptotics at fixed values of $t_r$ and $\varphi$
gives the final bound
\begin{equation}
|A(s,z)|_{\,s\to+\infty}<\left(\frac2{\pi\,\alpha_0}
\right)^{\frac3{\,2}}
\left(\sin^2\varphi+\frac{t_r}{t_0}\,\alpha_0^2\right)^{-
\frac1{\,4}}\frac{\sqrt{Y}\,~e^{\,Y\sqrt{t_r/t_0}}}
{\sqrt{t_r/t_0}\,(1-\sqrt{t_r/t_0})}\,\,.
\label{zbound3}
\end{equation}
If $\sin\varphi\neq0\,$, relation between the nonphysical
boundaries (\ref{tbound3}) and (\ref{zbound3}) for the fixed-$t$
and fixed-$(\varphi,\,t_r)$ cases is essentially the same as
between the physical boundaries (\ref{tbound1}) and
(\ref{nfbound1}) for the fixed-$t$ and fixed-$\theta_s$ cases
(recall that in the physical region $t_r$ equals zero and,
thus, is always fixed). If $\varphi\to0$ faster than $q_s^{-1}$,
then $t$ tends toward $t_r\,$; inequality (\ref{zbound3}) takes the
same form as the previous inequality (\ref{tbound3}) for
nonphysical positive $t$-values. If $\varphi\to\pm\pi\,$, also
tending toward the limit faster than $q_s^{-1}$, then
$u\to t_r\,$, and expression (\ref{zbound3}) comes to
correspond with the nonphysical fixed-$u$ case, for real
positive $u$-values.

\section{High energy behavior of amplitudes}

In the previous section we have constructed high-energy upper
boundaries for a $2\to2$ amplitude in different physical or
nonphysical configurations: forward or backward, fixed angle
scattering, fixed momentum transfer, and the nonphysical
interior of the Lehmann ellipse. In all the cases, the
boundaries have been expressed through two energy-dependent parameters, $\alpha_0$ and $Y$.

One of them, $\alpha_0$, has the clear and very simple
energy dependence. But it depends also on the parameter $t_0$,
related to the position of a crossed-channel singularity, which
we assume to be energy-independent. We have not discussed,
however, how the kind of the singularity could influence the
amplitude asymptotics.

The other parameter, $Y=\alpha_0 L$, depends on the number
$L$ of the partial-wave amplitudes that could be essential.
Dependence of $L$ (and $Y$) on energy is much less clear than
for $\alpha_0$. It is definitely not fixed by kinematics or
any physical principles.

Now we are going to discuss both problems in some detail.

\subsection{Role of different singularities}

Up to now we have assumed $t_0$ to be related with the position
of any $z$-singularity, nearest to the physical region.
At very high values of $l\,$, contribution of this singularity
is definitely the largest one. However, at $l$-values near
$L\,$, it may appear less essential than that of some more
remote, but more intensive singularity. Consider this point
more specificaily.

Let the nearest singularity be a pole corresponding to a
one-particle exchange, say, in the $t$-channel. If so, we can
present the amplitude as
$$ A(s,t)=\frac{r(s)}{t-t_0}+\tilde A(s,t)=
\frac1{2\,q^2_s}\cdot\frac{r(s)}{\,z-z_0}+\tilde A(s,z)\,,$$
where $\tilde A(s,z)$ has no $z$-singularities inside the
Lehmann ellipse related to $z_0\,$. Substituting this form
to Eq.(\ref{cint}), we separate the simple pole contribution
\begin{equation}
a_l^{(pole)}=-\frac{\pi\,r(s)}{2q_s\sqrt{s}\,}\cdot Q_l(z_0)\,,
\label{pole}
\end{equation}
while contribution of other singularities retains the form similar
to the contour integral (\ref{cint}), with $\alpha=\tilde\alpha>
\alpha_0$. The value of $\tilde\alpha$ may be increased up to
$\alpha_1$, corresponding to the next nearest singularity. Now,
applying again the asymptotic relation (\ref{asymq}), we can
rewrite boundary (\ref{boundam}) as
\begin{equation}
|a_l|<\frac{q_s}{\sqrt{s}}\cdot\frac{|r(s)|}{q_s\,\sqrt{s}}
\,\left(\frac{\pi}2\right)^{\frac32}
\cdot\frac1{\sqrt{e^{\alpha_0}\,
\sinh\alpha_0}}\cdot\frac{e^{-\alpha_0\,l}}{\sqrt{l}}
+\frac{q_s}{\sqrt{s}}\cdot\tilde B_1(s)\cdot\sqrt{e^{-\alpha_1}\,
\cosh\alpha_1}\cdot\frac{e^{-\alpha_1\,l}}{\sqrt{\,l}}\,,
\label{boundam1}
\end{equation}
where $\tilde B_1(s)$ is related with $\tilde A(s,z)\,$. At
very large $l\,$, the first term in the right-hand side is
always the leading one, since $\alpha_1>\alpha_0$. But this
right-hand side itself is then small and decreasing. The
situation may be different, however, at $l\approx L\,$, where
the right-hand side is near unity.

Let us consider the high-energy behavior of the boundary
(\ref{boundam1}). The asymptotics of the $t$-channel pole
residue $r(s)$ is $\sim s^J$ beeing directly related to the
spin $J$ of the corresponding hadron. There exist only two kinds
of hadrons which are stable under strong interactions and can,
thus, generate poles on the physical Riemann sheet. They are
basic pseudoscalar mesons with $J=0$ (pions, kaons, eta, and
so on) or basic baryons with $J=1/2\,$. Neither of the
corresponding exchanges can produce increasing total cross
sections. Moreover, their contribution to the boundary
(\ref{boundam1}) (see the first term) is vanishing when $s$
grows at fixed $l\,$. Increasing total cross sections could be
induced by exchanges of the higher-spin hadrons, but such
hadrons reveal themselves only as resonances. Therefore, the
corresponding poles are positioned at nonphysical Riemann
sheets and contribute to $\tilde B_1(s)$. Thus, to provide an
increasing cross section, the role of the second term should
grow at high $s$, due to growing $\tilde B_1(s)$.

If $\tilde B_1(s)$ grows indeed with $s$, the value of $L$ at
sufficiently high energy becomes determined by the second term
in the right-hand side (\ref{boundam1}), while the first term
becomes inessential. If the singularity at $\alpha=\alpha_1$ is
also a pole, we can separate it as well. After all, possible
increase of the total cross sections appears to be related
to the nearest threshold, even if there is a nearer pole.
Therefore, we can use formulas of the preceding sections
assuming that $\alpha_0$ always corresponds to the nearest
threshold and not to a pole. Note that the Lukaszuk-Martin
boundary~\cite{lukmart} for total cross sections, widely
discussed in the literature, uses just the nearest threshold,
which is the two-pion threshold.

\subsection{Energy dependence of boundaries}

Now we return to the explicit high-energy behavior of
parameters $\alpha_0$ and $Y\,$, which determine upper
boundaries for amplitudes in different scattering
configurations. At high energies,
\begin{equation}
\alpha_0\approx \frac{\sqrt{t_0}}{q_s}\approx
2\,\sqrt{\frac{t_0}{s}}\,,
\label{al0}
\end{equation}
where $t_0$, as explained, is related to the nearest threshold
in the crossed channel.

Less-evident energy dependence of the other parameter, $Y\,$, is
seen in Eq.(\ref{L1}), which can be rewritten as
\begin{equation}
e^{Y}\cdot\sqrt{Y}=\frac{B_0(s)\,\sqrt{\alpha_0}\,}2\,.
\label{Y}
\end{equation}
The quantity $B_0(s)$ is linearly related to the amplitude
integrated over nonphysical configurations, as determined
by Eq.(\ref{bint}) with $\alpha=\alpha_0\,$. Evidently, the
energy behavior of $Y$ (and, therefore, of boundaries for the
amplitude in physical configurations) directly depends on the
(unknown) energy behavior of $B_0(s)\,$.

The original Froissart boundaries have similar structure.
Instead of our parameter $Y$, those boundaries contain
\mbox{$Y_{\mathrm{Fr}}=\alpha_0\,L_{\mathrm{Fr}}\,$.} Recall
that in Froissart's notations $\alpha_0=\log(x_0+\sqrt{x_0^2-
1})\,$. The value of $L_{\mathrm{Fr}}\,$ at high energies was
chosen so that
$$e^{Y_{\mathrm{Fr}}}=B_{\mathrm{Fr}}(s)\,e^{N\alpha_0}\,,$$
where $N$ is the number of necessary subtractions (see
Ref.\cite{frois}, upper  right column on p.1055). Though both
$B_{\mathrm{Fr}}(s)$ and our $B_0(s)$, by construction, are
linearly related with the amplitude in nonphysical
configurations, those configurations are, generally, different
(they correspond to different values of $z$ or, equivalently,
$t$). Therefore, $B_{\mathrm{Fr}}(s)$ looks not identical to
$B_0(s)\,$. But if we assume an increasing total cross section,
both $B_{\mathrm{Fr}}(s)$ and $B_0(s)\,$ should increase as
well. It is reasonable to assume that they have similar
high-energy asymptotics. Then $Y_{\mathrm{Fr}}>Y$, and
Froissart's boundaries are higher than ours. In particular,
$$\sigma_{tot}<C\,Y^2<C\,Y_{\mathrm{Fr}}^2\,.$$

Discussion (and/or derivation) of any dispersion relation
always contains two ingredients. The main one is, of course,
knowledge of positions for singularities. According to standard
assumptions, the  character and position of a singularity is
determined by the unitarity condition. Then, the amplitude
singularities are only poles and branch-points, corresponding
to one-particle states or to several-particle thresholds
respectively. Their positions are determined, therefore, by
the particle masses. Any singularity of other kinds
(\textit{e.g.}, essential singularity) is not suggested by the
unitarity and, hence, is not expected to appear at some final
distance.

Unitarity, by itself, says nothing about analytical properties
of the infinite energy point. Nevertheless, a familiar
assumption is that this point has no essential singularity as
well. It means, in particular, that when $s\to\infty$ along any
direction at the physical Riemann sheet, the amplitude $A(s,t)$
can increase not faster than some limited power of $s\,$ (at
any fixed value of $t$ or $u$). This is the second important
ingredient, after which the dispersion relation in $s$ (with
some limited number of subtractions) arises just as a simple
manifestation of the Cauchy theorem.

If $A(s,t)$ is restricted by $s^n$, then this is true also for
both $B_{\mathrm{Fr}}(s)$ and $B_0(s)\,$. This implies that
$Y_{\mathrm{Fr}}$ grows no faster than $n\cdot\log s$, and
$\sigma_{tot}$ grows no faster than $\log^2 s$. It is just
the canonical formulation of the Froissart theorem.

Our Eq.(\ref{Y}) gives a more complicated, but somewhat
stronger, restriction for $Y$:
\begin{equation}
Y<\left(n-\frac14\right)\log s-\frac12 \log Y\approx
\left(n-\frac14\right)\log s-\frac12 \log\log s\,.
\label{Y1}
\end{equation}
Of course, it is smaller than the Froissart boundary $n\log{s}$.
If we describe the high-energy boundary for $Y$ by the standard
parametrization  $\sim\log(s/s_0)$, then Eq.(\ref{Y1})
means that the scale $s_0$ itself should be energy-dependent:
it should slowly (logarithmically) increase with energy.

Recall now that, according to Eq.(\ref{fbound2}),
the high-energy boundary for $\sigma_{\mathrm{tot}}$ is determined by
$Y^2$. Then we see that $\sigma_{\mathrm{tot}}$, indeed, cannot
grow faster than $\sim\log^2(s/s_0)$ with a fixed scale
$s_0\,$. But such growing log-squared behavior can be saturated
only if the scale $s_0$ used here is also increases with energy.

\section{Discussion of the results}

Let us summarize the above results for quantum amplitudes of
$\,2\to2$ processes.
\begin{itemize}

\item{Dispersion relations, single or double, are not
necessary for the Froissart theorem. Moreover, in the above
considerations we have not assumed any specific nature of the
underlying interaction. And even more, all the above relations
are consequences of the rather general quantum picture. They
could be equally applied either to the nonrelativistic
Schr\"odinger equation (using the energy $E$ instead of the
invariant $s$) or to relativistic interaction(s) of
(non-)elementary particles.}

\item{Unitarity is known since the original paper~\cite{frois}
to be a necessary input for the Froissart theorem. In all
cases, it restricts the partial-wave amplitudes, which
contribute to the total amplitude.}

\item{One more necessary input for the Froissart theorem is
absence of singularities in the physical region of $z=
\cos\theta$ (inside or on the edges of the interval $[-1,+1])$.
In the nonrelativistic case, this may be ensured by properties
of the potential (as, \textit{e.g.}, for the Yukawa potential).
In the relativistic case, this may be provided by the unitarity
condition in the crossed channel(s), if no massless exchanges
are possible (note the double-sided role of the unitarity in
the relativistic case). It is just the reason why the
Froissart theorem may be applied to strong interactions
(having finite-mass pions as the lightest particles), but not
to electroweak interactions (having the massless photon).
Absence of any physical-region singularities guarantees the
exponential smallness of high-$\,l$ partial-wave amplitudes.
As a result, only a finite number of partial waves may be
``essential'' at each given energy. }
\item{A very important ingredient of the Froissart theorem
comes from the mathematical properties of the Legendre
functions. Our calculations clearly demonstrate that high-energy
asymptotics for amplitudes in different configurations is
directly coupled with high-$l$ behavior of the Legendre
functions, which have $l=\infty$ as the essential singularity.
It is well-known that the Legendre polynomials $P_l(z)$
at large $l$ behave very differently inside the physical
region (the real interval $-1\leq z\leq+1$), at its edge
($z=\pm1$), and outside it. At the edge $|P_l(\pm1)|=1$, while
inside the region $|P_l(z)|<1$ and decreases $\sim l^{-1/2}$
with growing $l$. Outside the physical region, it exponentially
grows. Just these well-known facts imply that the quantum
amplitude has very different behavior inside the physical
region, outside it, or at the edge. They also explain why
high-energy asymptotics of the amplitude is much more moderate
in the physical region than outside it.}
\item{The most disputable input is given by assumptions on the
high-energy behavior of the amplitude in the nonphysical
region. The familiar assumption is the power boundary for the
increase in $s$, with some restricted power which is universal,
in the sense that it is applicable for any value of $t$ (or
$\cos\theta_s$), physical or nonphysical, and even complex.
The canonical log-squared bound for the total cross sections
arises as a consequence of such a restriction for the amplitude
at nonphysical values of $t$. }
\end{itemize}

Let us consider in some more detail the problem of
discontinuities between asymptotics for different
configurations. Froissart's calculations~\cite{frois} show
different high-energy behavior for $\theta_s=0$ and \mbox{
$\theta_s\neq0$}. Our calculations confirm this result and
present it also as discontinuities of asymptotics between $t<0$,
$t=0$, and $t>0$. At first sight, this looks strange since
$t=0$ is a non-singular point, where the amplitude is analytic
(and continuous). Therefore, it would be natural to expect the
$s$-asymptotics to be continuous as well. However, the Legendre
functions clearly demonstrate just the opposite behavior. The
point $z=1$ is always an analyticity point for $P_l(z)$.
Nevertheless, the high-$l$ asymptotics is discontinuous near
$z=1$. Indeed, at the real axis below this point
$$P_l(\cos{\theta})\approx\sqrt{\frac2{\pi l\,
\sin{\theta}}}\,\cdot\,\cos\left[\left(l+
\frac12\right)\,\theta-\frac{\pi}4\,\right];$$
at the point itself
$$P_l(1)=1\,;$$
at last, above this point,
$$P_l(\cosh\beta)\approx\frac1{\sqrt{2\pi\,l}}\,\cdot
\frac{e^{\beta(l+\frac1{\,2})}}{\sqrt{\sinh\beta}}\,\,,
~~~~\mathrm{Re}\,\beta>0\,.$$
Discontinuities of the asymptotics are evident here. However,
they appear only in the limit $l\to\infty$ at fixed $z$. If one
takes $l$ to be large but fixed, $P_l(z)\,$ is, of course,
continuous in $z$, as seen from one more well-known approximate
relation (\ref{pbes}). It shows that the different asymptotics
join in a narrow intervals $(z-1) \sim l^{-2}$.

A similar conclusion is true for the amplitude as well.
Equation (\ref{tbound}) is continuous near $t=0$, if $Y$ is finite.
But essential changes of the right-hand side take place in a
very narrow interval $\Delta t\sim t_0/Y^2\,$. In the limit
$Y\to\infty$ we obtain discontinuous boundaries for different
values of $t$ near zero. Analogous is the transition from
physical real values of $t<0$ to nearby complex values of $t$.

Now we briefly discuss the problem of power high-energy
behavior for the amplitude. In quantum field theory, such
behavior could not be deduced from any general principles
(in particular, unitarity can say nothing on this problem).
The only motivation for the power behavior near infinity is
that it allows us to write dispersion relations.

In difference, for quantum mechanics, the high-energy
asymptotics of the amplitude can be found somehow, if the
potential is given. Dispersion relation in energy is true for
the forward scattering amplitude with many quantum-mechanical
potentials~\cite{lanlif}. Potentials which admit dispersion
relations for nonforward scattering seem to be much rarer.
But at least for Yukawa-like potentials, the amplitude
satisfies even the Mandelstam representation, which is the
double-dispersion relation in energy and momentum
transfer~\cite{regge}.

In the relativistic case, the problem of high-energy
asymptotics for two-particle amplitudes is still open. Even the
single-dispersion relation in energy has been mathematically
proven only for the pion-nucleon elastic scattering in the
forward direction or in some finite interval of real negative
(\textit{i.e.}, physical) values of $t$~\cite{BMP}. Note,
however, that this does not prove the Froissart log-squared
behavior of the pion-nucleon $\sigma_{\mathrm{tot}}$. For the
forward dispersion relation to be true, $\sigma_{\mathrm{tot}}$
may grow faster than the canonical Froissart bound, though not
faster than some finite power of $s$.

For better understanding of the situation, it is interesting to
look for hints from the perturbation theory. Summation of the
Feynman diagrams was most intensively investigated for QED and
perturbative QCD (pQCD). In both cases, sums of essential
logarithms for ``one-tower'' diagrams provide the power
behavior of the high-energy asymptotics for
$\sigma_{\mathrm{tot}}$. The corresponding exponents are small:
$\sim\alpha^2$ in QED~\cite{ChWu1, ChWu2, GLF} and
$\sim\alpha_s$ in pQCD~\cite{BFKL}. The difference is due to
different forms of interactions between the corresponding gauge
bosons: through electron loop(s) for the photons in QED, and through gluon
exchange(s) for the gluons in pQCD.

The authors of ref.\cite{ChWu2} believe that ``unitarity is
violated'' because of such power behavior. Therefore they
consider it to be transient. As they hope, it will be changed by
the log-squared behavior after summing up all ``multitower''
diagrams, though the authors agree that ``this method has no mathematical justification''.

We have seen, however, that violation of the log-squared
asymptotics is not necessarily related to violation of unitarity.
It may mean violation of power bounds for nonphysical amplitudes.
On the Lehmann ellipse, the values of momentum transfers may
reach large complex values $|t|\approx|u|\approx s/2,~
|\mathrm{Im}t|\approx|\mathrm{Im}u|\approx\sqrt{st_r}\,$, with
$t_r$ being fixed at the given Lehmann ellipse. High-energy
behavior for amplitudes in such configurations has never been
investigated.

There is one more reason why cross section estimates in QED
and/or pQCD might be doubtful. Both theories provide massless
(photon/gluon) exchanges, which generate $z$-singularity at
the edge of the physical region. For scattering of charged
(colored) objects this singularity is nonintegrable (it is the
pole due to one-photon/one-gluon exchange) and makes the total
cross section infinite at any energy. Thus, discussion of any
bounds for $\sigma_{\mathrm{tot}}$ becomes meaningless.

However, for neutral (colorless) objects, the corresponding
singularities, though being also on the edge of the physical
region, are integrable and do not provide permanently infinite
cross section. In such situation, the nearest singularity(ies),
just at the edge of the physical region, may appear less
essential for the asymptotics than more distant singularities
(we have seen above how the nearest pole could be inessential
at high energies, as compared with more distant contributions).
Then, the Froissart approach might be applicable for neutral
(colorless) objects in QED (pQCD), even despite the possibility
of massless exchanges. This needs, however,  special
investigation. If confirmed, results of diagram summation
for QED and pQCD, briefly described above, could give indeed
serious theoretical hints for power (though rather slow)
increase of hadron cross sections.

Strong interaction phenomenology presents also some other
evidences, though indirect, for power increase of the total
cross sections. For example, an essential input to prove the
power asymptotics (and the Mandelstam representation) of
amplitudes for nonrelativistic scattering in Yukawa-type
potentials was a restricted value of real parts for Regge
trajectories in such potentials~\cite{regge}. The
phenomenological evidence for linearity of hadron Regge
trajectories, if true, means that the relativistic amplitudes,
at least in some configurations, may grow faster than any power
of energy. Correspondingly, the total cross sections may grow
faster than $(\log{s})^2\,$.

In summary, the real content of the Froissart theorem is the
much softer high-energy behavior of physical amplitudes (and
total cross sections) as compared  to behavior of nonphysical
amplitudes. This is implied by the physical requirements of
unitarity and absence of massless exchanges, together with
the mathematical properties of the Legendre functions.
Dispersion relations, either in energy or in momentum
transfers, are not necessary. Moreover, the nature of
interaction is, by itself, inessential; however, strong
interactions are marked out by the absence of massless
particles.

The specific form of the high-energy bound for amplitudes in
physical configurations (and, thus, for the total cross section
as well) is directly correlated with the high-energy behavior
for amplitudes in nonphysical configurations (in particular, at
large and complex $t$). The canonical log-squared bound
corresponds to power asymptotics of the nonphysical amplitudes
(which can be ``hidden'' in dispersion relations with a finite
number of subtractions). Its violation, contrary to folklore in
the literature, would not mean violation of unitarity. It may
mean only that the amplitude in nonphysical configurations can
grow with energy faster than any power of $s$. Such possibility
does not seem to contradict any basic principles. Precise
measurements of cross sections at very high energies (at LHC,
in particular) can possibly help to discriminate between
logarithmic and/or power asymptotics. Other high-energy
observables may also be helpful.

\section*{Acknowledgments}
The author thanks D.~I.~Diakonov and L.~N.~Lipatov
for stimulating discussions. The work was partly
supported by the Russian State grant RSGSS-65751.2010.2.

\section*{APPENDIX. Sums and integrals for boundaries}

When constructing boundaries for amplitudes in different
configurations, we need to calculate the high-energy behavior
for sums of the form
$$ S=\sum_{l_1}^{l_2}\,l^k\cdot f(\alpha l)\,,$$
where summation runs on integer values of $l$, and $\alpha\to0$
at $s\to+\infty$. Let us rewrite the sum as
$$ S=\sum_{l_1}^{l_2}\,l^k\cdot f(\alpha l)\cdot\Delta l\,,$$
where $\Delta l=1$. Now, define the new variable $y=\alpha l$.
Then the sum takes the form
$$ S={\alpha}^{-(k+1)}\,\sum_{y_1}^{y_2}\,y^k\cdot f(y)\cdot
\Delta y\,,$$
with summation running on the $y$-points, corresponding to the
integer values of $l$, with intervals $\Delta y=\alpha\,$.
When $s\to+\infty$ (\textit{i.e.}, $\alpha\to0$), the latter
sum tends toward the integral
$$ S\approx{\alpha}^{-(k+1)}\,\int_{y_1}^{\,y_2}\,y^k\cdot
f(y)\cdot d y\,.$$
For all our boundaries we have used just such integrals. As an
illustration, let us consider the simple sum
$$ \sum_{0}^{L-1}\,(2l+1)=L^2\,.$$
The above procedure, with $Y=\alpha L$, transforms it into the
sum of two integrals
$$ {\alpha}^{-2}\int_0^{(Y-\alpha)}\,2y\,dy+{\alpha}^{-1}
\int_0^{(Y-\alpha)}\,dy=\left(\frac{Y-\alpha}{\alpha}
\right)^2+\left(\frac{Y-\alpha}{\alpha}\right)=\frac{Y(Y-\alpha)}
{\alpha^2}=L^2\left(1-\frac1{L}\right)\,.$$
If $L$ is growing at $s\to+\infty$, the main term is correctly
reproduced.
$$~~~~~~~$$
$$~~~~~~~~$$

\end{document}